\documentclass[letterpaper,twocolumn,fleqn]{article} 

\usepackage[pass,letterpaper]{geometry} % Afin d'avoir A US-Letter-sized PDF (demandé par EI).

\usepackage{ist}
\usepackage{epsfig}
\usepackage{epstopdf}  % gestion des eps
\usepackage{amssymb}   % Pour les polices Math
\usepackage{hyperref}   % Pour les lien url clickables
\usepackage{amsmath}    % Pour affichage matrice
\usepackage{enumitem}   % pour les énumération avec des nombres romains
\usepackage[table,xcdraw]{xcolor}
\usepackage{diagbox}
\usepackage{stfloats}
\title{
\vspace{-2cm}
{\footnotesize \textnormal{\textit{Paper accepted to Media Watermarking, Security, and Forensics, IS\&T Int. Symp. on Electronic Imaging, SF, California, USA, 28 Jan. - 2 Feb. 2018. \vspace{-0.4cm}\\}}
{\tiny \textnormal{\textit{Last minor modifications made the 3rd of February 2018.\\}}}
}\vspace{+0.2cm}\\ How to augment a small learning set for improving the performances of a CNN-based steganalyzer?}
\author{Mehdi YEDROUDJ${}^{1,2}$, Marc CHAUMONT ${}^{1,3}$, Fr\'ed\'eric COMBY${}^{1,2}$ \\
\small LIRMM${}^1$, Univ Montpellier${}^2$, CNRS, Univ N\^imes${}^3$, Montpellier, France\\
\small\texttt{\small\{mehdi.yedroudj, marc.chaumont,frederic.comby\}@lirmm.fr}}

\date{} % date has an empty field.

\begin{document} 

\maketitle 

\thispagestyle{empty} % prevents the first page to be numbered

%\keyword{Steganalysis, Deep Learning, Convolutional Neural Network (CNN), Base Augmentation, Clairvoyant Scenario, Cover-Source Mismatch.}

%%%%%%%%%%%%%%%%%%%%%%%%%%%%%%%%%%
% Abstract
%%%%%%%%%%%%%%%%%%%%%%%%%%%%%%%%%%

\begin{abstract}

Deep learning and convolutional neural networks (CNN) have been intensively used in many image processing topics during last years. As far as steganalysis is concerned, the use of CNN allows reaching the state-of-the-art results. The performances of such networks often rely on the size of their learning database. An obvious preliminary assumption could be considering that “the bigger a database is, the better the results are”. However, it appears that cautions have to be taken when increasing the database size if one desire to improve the classification accuracy i.e. enhance the steganalysis efficiency. To our knowledge, no study has been performed on the enrichment impact of a learning database on the steganalysis performance. What kind of images can be added to the initial learning set? What are the sensitive criteria: the camera models used for acquiring the images, the treatments applied to the images, the cameras proportions in the database, etc? This article continues the work carried out in a previous paper in submission \cite{Yedroudj2018_Net}, and explores the ways to improve the performances of CNN. It aims at studying the effects of “base augmentation” on the performance of steganalysis using a CNN. We present the results of this study using various experimental protocols and various databases to define the good practices in base augmentation for steganalysis.

\end{abstract}

%%%%%%%%%%%%%%%%%%
% Overall Document 
%%%%%%%%%%%%%%%%%%
\section{Introduction}
\label{sec:intro}

Convolutional neural networks (CNN) became very popular to solve classification problems in the last five years. Several authors have proposed to use CNNs to solve steganalysis problems \cite{Qian_2015_Deep}, \cite{Pibre2016}, \cite{Xu2016a}, \cite{Ye2017}. These methods yield encouraging results but remained comparable to the state-of-the-art algorithms performances. Authors have explored many approaches to improve it such as using a phase split \cite{Chen2017}, an ensemble of CNN \cite{Xu2016b}, the transfer learning \cite{Qian2016_Transfer} or the augmentation of the database \cite{Ye2017}, \cite{Zeng2017_Millions}. 

Let us put aside the quest of the best deep learning network architecture for the steganalysis task. In this paper, our objective is to look at a "real-world" problem \cite{Ker2013_RealWorld}, which is to learn with a small size database. This problem is also known as low regime learning. It is well-known that supervised approaches based on the use of CNNs need a lot of samples when used for steganalysis purposes. The seminal propositions of Qian {et al.} \cite{Qian_2015_Deep} and Pibre {et al.} \cite{Pibre2016} used from 8 000 to 80 000 spatial images resized to 256$\times$256 (BOSSBase \cite{Bas2011-BOSS} or ImageNet \cite{Krizhevsky_AlexNet_2012}). In 2017 the authors mainly use around 5 000 pairs of images \cite{Xu2016a}, \cite{Ye2017}, \cite{Chen2017}, \cite{Xu2017}, which is probably insufficient. The number of images for the learning has even reached five millions of samples in \cite{Zeng2017_Millions}. 
% YeNet: 10 000 de base. Regarder les autres...
% 512x512 Xu-Net : 10 000 (cover+stego)
% JPEG-Phase Fridrich 60%*10 000 *2 = 12 000
% Detect J-UNIWARD : 10 000

In an operational and realistic protocol, the number of available images for the learning task could be much smaller than what is used in "laboratory". Because all the CNN-based steganalysis are sensitive to the cover-source mismatch phenomenon \cite{Cancelli2008, Ker2014_Mishmash, kodovsky2014study}, each time the source distribution is modified, the learning process has to be restarted. The aim of this paper is thus to look at the impact of {\it artificial data-augmentation}, which is probably more realist than having access to a huge database of a given source distribution. In all cases, using data-augmentation is an automatic process which requires less human time consumption than searching for images of similar distributions. %Poursuivre l'argumentation en faveur de l'étude sur l'augmentation automatique de la base : pour les GAN : mettre ref Ian Goodfellow + Jiwu. Argumentation également sur la compréhension du phénomène de cover-source mismatch. 

Today, the classical scenario used to test an embedding algorithm efficiency is to use the BOSSBase \cite{Bas2011-BOSS} for training and testing, assigning 5000 of the 10000 images to the learning database, while the rest used as testing database. A classical way to artificially increase the learning database without changing the labels is to flip and rotate the learning database without interpolation \cite{Krizhevsky_AlexNet_2012}. 

Recently, Ye {\it et al.} \cite{Ye2017} proposed to increase the size of the training database, by adding to the initial 50\% of BOSSBase, the whole BOWS2 \cite{BOWS2008} database (this gives a total of 15000 pairs of images for the training set), while the test set is unchanged and is made of the remaining 50\% of BOSSBase. This process effectively improves the results in terms of error probability of detection. However, it could be considered as {\it a very lucky measure} because the improvement is essentially due to the fact that BOSSBase and BOWS2 share some identical camera models, and a similar "development" process\footnote{The "development" stands for the numerical processes transforming a color RAW image to a 256$\times$256 8-bit grey-levels image}. 

The question is thus still open: how should we process in order to enrich a learning database? Can we enrich even more the BOSS learning base in order to obtain a huge learning base, and thus improve the steganalysis results? In this paper we intend to  experimentally explore efficient ways to {\bf increase} the learning database of a CNN based steganalyzer. In Section \ref{sec:cnn}, we recall the topology of the CNN used for the various experiments \cite{Yedroudj2018_Net}. In Section \ref{sec:methodo}, we describe the experimental protocol and briefly present all the setups. In Section \ref{sec:exp}, we experimentally explore the different augmentation methods and we draw conclusions on the practical question of the learning database augmentation.

%%%%%%%%%%%%%%%%%%%%%%%%%%%%%%%%%%%%%%%%%%%%%%%%%%%%%%%%%%%%%%%%%%%%%%%%%%%%%%%%
\section{Yedroudj-Net CNN}
\label{sec:cnn}

In this paper, our study on the data augmentation for spatial steganalysis is conducted only on the Yedroudj-Net \cite{Yedroudj2018_Net}. This CNN has been created in 2017 and is a mix of the Xu-Net \cite{Xu2016a} and Ye-Net \cite{Ye2017}, which are the two best CNNs created up to 2017 for steganalysis purposes. Yedroudj-Net gives better results than Xu-Net \cite{Xu2016a} and Ye-Net \cite{Ye2017} on WOW \cite{Holub2012_WOW} and S-UNIWARD \cite{Holub2014}, and also provides better results than an Ensemble Classifier \cite{Kodovsky2012-EnsembleClassifiers} with a Rich Model \cite{Fridrich2012_Rich} when compared on a baseline where there is only one CNN, and no tricks such as the use of an ensemble or transfer learning. We have also conducted database augmentation experiments on Xu-Net \cite{Xu2016a} and Ye-Net \cite{Ye2017} and they follow the same trend as Yedroudj-Net. 

\begin{figure*}[t]
\centering
\includegraphics[width=18cm,height=4cm]{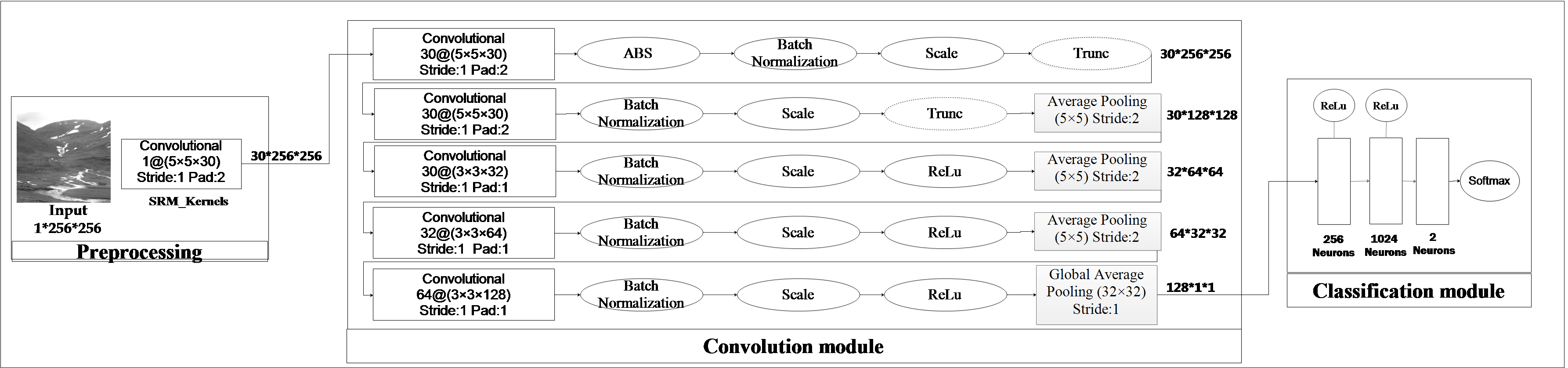} 
\caption{Yedroudj-Net CNN architecture. Figure taken from \cite{Yedroudj2018_Net}.}
\label{fig:yedroudj-net}
\end{figure*}

Yedroudj-Net is composed of a {\it pre-processing block}, five {\it convolutional blocks}, and a {\it fully connected block} made of three fully connected layers followed by a {\it softmax}. The network produces a probability distribution over the two class labels: stego or cover image. Fig. \ref{fig:yedroudj-net} illustrates the overall architecture of our CNN.

For more details on Yedroudj-Net, the reader can have a look at the paper \cite{Yedroudj2018_Net} and the online code at \url{http://www.lirmm.fr/~chaumont/DemoAndSources.html}
%http://www.lirmm.fr/ $\tilde{ }$ chaumont/DemoAndSources.html. 
Note that the hyper-parameters are kept identical. 

%%%%%%%%%%%%%%%%%%%%%%%%%%%%%%%%%%%%%%%%%%%%%%%%%%%%%%%%%%%%%%%%%%%%%%%%%%%%%%%%
\section{Experimental methodology}
\label{sec:methodo}

\subsection{Objectives and Dataset baseline}
\label{sec::setup}

Our final objective is to {\bf increase the size of the learning database} of a CNN based steganalysis through data-augmentation in order to improve its performances. Indeed, increasing the number of learning samples is often beneficial for learning efficient features dedicated to a specific task. But, for steganalysis, the samples have to be selected carefully. The "new" samples have to share a "similar distribution" compared to the "original" samples. One thus tries to find {\bf distribution-preserving transformations} which, when applied on an input cover or precover image, generate synthesized images that follow the same distribution. Those synthesized images could then be integrated into the learning database as additional images in order to increase the CNN classifier efficiency.

In this paper, first, we explore the factors that are influencing a cover distribution such as the camera model, or the development, and second, we propose {\it distribution-preserving} transformations that allow to enrich an initial database and to improve the CNN efficiency.

Our baseline setup will thus be working with the BOSSBase split into two sets. We assign 50\% of the cover/stego pairs to the "original" training set, and the rest, to the testing set. For the training set, 4000 out of 5000 pairs are randomly selected for training and the remaining 1000 pairs are set aside for validation. Thus, the testing set is made of 5000 pairs. {\bf Regardless of the learning database enrichment, the test database will always contain images from and only from BOSSBase}. For a fair comparison, we will use the same test base for all the experiments. To summarize, the learning set will always contain at least 4000 pairs of BOSSBase images, and the {\bf validation set will always contain 1000 pairs of BOSSBase images}.

\subsection{Software platform}
We used S-UNIWARD \cite{Holub2014}, and WOW \cite{Holub2012_WOW}, two well-known content-adaptive methods for the embedding in the spatial domain. Note that we used the Matlab implementations (online codes\footnote{http://dde.binghamton.edu/download/}) with the simulator for the embedding and a random key for each embedding. We thus avoid any wrong use of the C++ codes, i.e. a fixed and unique embedding key, as reported in \cite{Pibre2016}.

All experiments were performed with the publicly available {\it Caffe} toolbox \cite{caffe_jia} with necessary modifications, plus digits V5. All tests were run on an NVidia Titan X GPU card.

\subsection{Datasets}
\label{Training_section}

Due to our GPU computing platform and time limitation, we conduct all the experiments on images of 256$\times$256 pixels. To this end, we resampled all the 512$\times$512 images to 256$\times$256 images, using the {\it imresize()} Matlab function with the default parameters (bicubic interpolation with anti-aliasing).

For the various experimental setup, we are using the different databases listed below, and convert them to $256x256$ images:
\begin{itemize}
\item the BOSSBase v1.01 \cite{Bas2011-BOSS} consisting of 10 000 grey-level images of size $512\times512$, never compressed, and coming from 7 different cameras, 
\item the BOWS2 \cite{BOWS2008} consisting of 10 000 grey-level images of size $512\times512$, never compressed, and whose distribution is close to BOSSBase,
\item the LIRMMBase \cite{Pibre2016} consisting of 9 388 grey-level images of size $512\times512$, never compressed, and coming from 7 different cameras. All the used cameras are different from those used in BOSSBase. This database is a variant of the LIRMMBase (\url{http://www.lirmm.fr/~chaumont/LIRMMBase.html}) where images with no semantic content have been suppressed. Note that the development (of the RAW images) used in order to obtain the 256$\times$256 images have been done reusing the same script than the one used for generating BOSS and BOWS2 (\url{http://www.lirmm.fr/~chaumont/LIRMMBase/macroProductPGM.sh}).
\item the PLACES2 \cite{zhou2017places} containing more than one million of JPEG images coming from unknown cameras. For the experiments, those images are decompressed, converted in grey-level images, and then resized.
\end{itemize}

For some experiments, we re-run a {\it development} process and we will use the {\it ImageMagick} free and open-source software.

During the CNNs training, we regularly observe the {\it Loss} and {\it Accuracy} curves, computed on the validation test, to manually stop the training when an over-fitting phenomenon appears. This over-fitting occurs when the {\it Loss} curve continues to decrease on the training set but starts to increase on the validation set. For all the experiments, we report the error probability evaluated on the testing set.

\subsection{Description of the different experimental setups}

Below, we briefly listed all the experimental setups with a small description explaining each choice:
\begin{itemize}
\item {\bf Setup 1: Classical enrichment}. In this setup, the goal is to obtain the performance baseline. The enrichment of the {\it original} learning database (made of 4000 pairs) is obtained thanks to the virtual augmentation using the label-preserving flipping and rotations \cite{Ye2017}, and the enrichment with BOWS2  images. This experiment is presented in Section \ref{sec:classical_enrichment},

\item {\bf Setup 2: Enrichment with other cameras}. In this setup, the goal is to evaluate the gain/loss of adding images from different cameras from the ones used in the {\it original} learning set. This experiment is presented in Section \ref{sec:other_cameras},
% Faut il dire plus sur les cameras ?

\item {\bf Setup 3: Enrichment with strongly dissimilar sources and unbalance proportions}. In this setup, the goal is to evaluate the gain/loss of adding a huge number of images generated using cameras and a development, totally different from those used in the {\it original} learning set. This experiment is presented in Section \ref{sec:different_sources_unbalance},

\item {\bf Setup 4: Enrichment with the same RAW images but with a different development}. In this setup, the idea is to evaluate the gain/loss of adding the same {original} RAW images whose development is different from the one used for the {\it original} learning set. This experiment is presented in Section \ref{sec:RAWimages_different_dev},

\item {\bf Setup 5: Enrichment with a re-development of the learning set}. In this setup, the objective is to evaluate the gain/loss of adding the same {original} images which are {\it re-developed}. This experiment is presented in Section \ref{sec:images_redevelop},

\end{itemize}

%%%%%%%%%%%%%%%%%%%%%%%%%%%%%%%%%%%%%%%%%%%%%%%%%%%%%%%%%%%%%%%%%%%%%%%%%%%%%%%%
\section{Results and discussions}
\label{sec:exp}

%%%%%%%%%%%%%%%%%%%%%%%%%%%%%%%%%%%%%%%%%%%%%%%%%%%%%%%%%%%%%%%%%%%%%%%%%%%%%%%%

%%%%%%%%%%
%FRED 20/12/17
%Pour moi on pourrait mettre la description des SETUPs dans la partie expérimentation. Sinon cela créé des redites et des lourdeurs (la partie ou il faut se taper tous les setups est trop vague : on ne sait pas quelles sont les bases d'images utilisées, quels sont les developpements utilisés, etc...
%Sur le premier test je ne vois pas pourquoi on se compare avec le SRM + EC ? (tab 1) Ce résultat n'est utilisé nul part ailleurs ? en plus le résultat de YedroudjNet est ré-utilisé dans le tableau 2. A mon avis ce dernier est le seul utile.
% Pour moi on met en début de chaque expérience l'objectif visé, les mesures, les commentaires des résultats et on passe au suivant.
%%%%%%%%

\subsection{Setup 1: Classical enrichment}
\label{sec:classical_enrichment}

\begin{table}[htb]
\renewcommand{\arraystretch}{1.6}

\renewcommand{\arraystretch}{1.1}
\scalebox{0.8}{
\begin{tabular} {l|c|c|c|c|} 
\cline{2-3}
                                                                                 & \multicolumn{2}{c|}{\cellcolor[HTML]{656565}{\color[HTML]{FFFFFF} \textit{\textbf{BOSS 256$\times$256}}}}                           \\ \cline{2-3} 
\hline                                                                                 
\multicolumn{1}{|l|}{{\cellcolor[HTML]{B4ACAC}\backslashbox {Steganalysis}{Payload}}}     
& {\cellcolor[HTML]{C0C0C0}\color[HTML]{FFFFFF} \textbf{WOW 0.2 bpp}} & {\cellcolor[HTML]{C0C0C0}\color[HTML]{FFFFFF} \textbf{S-UNIWARD 0.2 bpp}} \\ \hline \cline{1-3}
\hline
\multicolumn{1}{|l|}{\cellcolor[HTML]{C0C0C0}{\color[HTML]{FFFFFF} Yedroudj-Net }} & {\bf 27.8} \%                            & {\bf 36.7} \%   \\ \hline \cline{1-3}
\multicolumn{1}{|l|}{\cellcolor[HTML]{C0C0C0}{\color[HTML]{FFFFFF} SRM+EC \cite{Fridrich2012_Rich,Kodovsky2012-EnsembleClassifiers}}}       & 36.5 \%                              & {\bf 36.6} \%  
\\ \hline \cline{1-3}
\end{tabular}}
\vspace{+0,4cm}
\caption{
Table 1: Steganalysis error probability of Yedroudj-Net, and SRM+EC for two embedding algorithms WOW and S-UNIWARD at 0.2 bpp and 0.4 bpp.\label{tab:Classical}}
\vspace{+0,5cm}
\end{table}

In Table \ref{tab:Classical}, we report the error probability obtained when there is {\bf no enrichment} which means there are 4000 pairs in the learning set (+ 1000 pairs in the validation set), and 5000 pairs in the test set. All images are from BOSSBase. For a cursory comparison, the performance is reported for the Yedroudj-Net, and the Spatial Rich Model + the Ensemble Classifier {\it (SRM + EC)}, for the embedding algorithm WOW \cite{Holub2012_WOW} and S-UNIWARD \cite{Holub2014} at payload 0.2 bpp.

Yedroudj-Net has an error probability 8\% lower for WOW algorithm at 0.2 bpp, and a similar error probability for S-UNIWARD  at 0.2 bpp compared to SRM+EC. As reported in \cite{Yedroudj2018_Net}, Yedroudj-Net obtains similar or better results compared to the state-of-the-art (including versus Xu-Net and Ye-Net) in a fair comparison setup.

%FRED 20/12/17
%A mon avis ce paragraphe ne sert à rien, en plus on ne donne pas les résultats avec Xu-net et Ye-net dans le tableau donc il ne faut pas en parler.

\begin{table}[htb]
\centering
\renewcommand{\arraystretch}{1.1}
\scalebox{0.9}{
\begin{tabular}{l|l|l|l|l|}
\cline{2-3}                                             
 & \cellcolor[HTML]{C0C0C0}WOW 0.2 bpp & \cellcolor[HTML]{C0C0C0}S-UNIWARD 0.2 bpp      \\ \hline                                     
\multicolumn{1}{|l|}{\cellcolor[HTML]{C0C0C0}BOSS}   & {\bf 27.8} \%   &  {\bf 36.6} \%                                                         \\ \hline
\multicolumn{1}{|l|}{\cellcolor[HTML]{C0C0C0}BOSS+VA} & 24.2 \%      & 34.8 \%  
\\ \hline
\multicolumn{1}{|l|}{\cellcolor[HTML]{C0C0C0}BOSS+BOWS2} & 23.7 \%        & 34.4\%                                         \\ \hline
\multicolumn{1}{|l|}{\cellcolor[HTML]{C0C0C0}BOSS+BOWS2+VA} & 20.8 \%      &31.1 \%  
\\ \hline
\end{tabular}}
\vspace{+0,4cm}
\caption{Table 2: Base Augmentation influence: error probability of  Yedroudj-Net, on WOW and S-UNIWARD at 0.2 bpp with and without Data Augmentation.\label{tab:base_augmentation}}
\vspace{+0,5cm}
\end{table}

In Table \ref{tab:base_augmentation}, we report the results with {no enrichment} (noted {\bf BOSS}), the results with the Virtual Augmentation (VA) of the BOSS's training set (noted {\bf BOSS + VA}; Virtual Augmentation consists in label-preserving flipping and rotations), the results with BOWS2 enrichment (noted {\bf BOSS + BOWS2}), and the results with BOWS2 enrichment + the Virtual Augmentation (noted {\bf BOSS+BOWS2+VA}). Some of these results have already been given in \cite{Yedroudj2018_Net}, are re-presented in order to have a self-containing paper. Note that for BOSS+BOWS2, the training set is made of 14 000 pairs (without counting the validation), 32 000 pairs for BOSS+VA (without counting the validation), and for BOSS+BOWS2+VA, the training set is made of 112 000 pairs (without counting the validation).

When the enrichment is obtained by only applying a virtual augmentation (BOSS+VA), a significant improvement is observed. The decrease of the error probability detection is 3\% for WOW (resp. 2\% for S-UNIWARD). This enrichment measure was initially proposed in \cite{Krizhevsky_AlexNet_2012} and it is indeed very efficient. The reader should understand that the VA is an easy and low-cost measure in order to significantly improve the performances.

One can also observe better performance when using BOSS+BOWS2 compared to only using BOSSBase. The CNN decreases its detection error probability by 4\% for WOW (resp. 2\% for S-UNIWARD). As stated in the introduction, BOSSBase and BOWS2 share some identical camera models and a similar "development" process. As also observed in Section \ref{sec:RAWimages_different_dev}, in a close setup, this enrichment setup ("similar cameras" + "similar development") allows to increase the performances. We guess that in that case, the added images increase the generalization capability of the network. 

When the enrichment is obtained with BOSS+BOWS2+VA, again a significant improvement is observed. The decrease of the error probability detection is 7\% for WOW (resp. 5\% for S-UNIWARD) compared to the no-enrichment setup. Note that the results given in the current Section will be the reference performances for the comparisons given in the next sections.

The observations given in this Section are confirming that if the database augmentation ensures a good diversity of the database, the CNN can improve its detection accuracy. The experiments described in the next sections are thus done in order to better understand the properties that have to be kept when adding images to the {\it original} database.

% \begin{figure}[t]
% \centering
% \includegraphics[width=9cm,height=7cm]{bows.png} 
% \caption{Imaginary vision of BOSS and BOSS2 images  distribution on a Hyperplane..}
% \label{fig:yedroudj-net}
% \end{figure}

%%%%%%%%%%%%%%%%%%%%%%%%%%%%%%%%%%%%%%%%%%%%%%%%%%%%%%%%%%%%%%%%%%%%%%%%%%%%%%%%
\subsection{Setup 2: Enrichment with other cameras}
\label{sec:other_cameras}

\begin{table}[htb]
\centering
\renewcommand{\arraystretch}{1.6}
\scalebox{0.8}{
\begin{tabular}{l|l|l|l|l|l|}
\cline{2-3}
                                                    
 & \cellcolor[HTML]{C0C0C0}WOW 0.2 bpp 
& \cellcolor[HTML]{C0C0C0}S-UNIWARD 0.2 bpp      \\ \hline                                     
\multicolumn{1}{|l|}{\cellcolor[HTML]{C0C0C0}BOSS}   & {\bf 27.8} \%                        & {\bf 36.7} \%                                                         \\ \hline
\multicolumn{1}{|l|}{\cellcolor[HTML]{C0C0C0}BOSS+LIRMM} & 29.9 \%                        & 38.6 \%                                         \\ \hline
\multicolumn{1}{|l|}{\cellcolor[HTML]{C0C0C0}BOSS+LIRMM+BOWS2} & 26.8 \%                        & 36.9 \%                                                         \\ \hline
\multicolumn{1}{|l|}{\cellcolor[HTML]{C0C0C0}BOSS+LIRMM+BOWS2+VA} & 25.7 \%                        & 36.1 \%                                                                     \\ \hline

\end{tabular}}
\vspace{+0,4cm}
\caption{Table 3: Base Augmentation influence: error probability of Yedroudj-Net, on WOW and S-UNIWARD at 0.2 bpp with a learning base augmented with either LIRMM, LIRMM+BOWS2, or LIRMM+BOWS2+VA. \label{tab:other_cameras}}
\vspace{+0,5cm}
\end{table}

In Table \ref{tab:other_cameras}, we report the results with {\it no enrichment} (noted {\bf BOSS}), the results with LIRMM enrichment (noted {\bf BOSS + LIRMM}), the results with LIRMM and BOWS2 enrichment (noted {\bf BOSS + LIRMM + BOWS2}), and the results with LIRMM and BOWS2 enrichment  + the Virtual Augmentation (noted {\bf BOSS + LIRMM + BOWS2 + VA}). Note that for {\it BOSS+LIRMM}, the training set is made of 14 000 pairs, for {\it BOSS+LIRMM+BOWS2}, the training set is made of 23 388 pairs (without counting the validation), and for the {\it BOSS+LIRMM+BOWS2}, the training set is made of 187 104 pairs (without counting the validation).

% \begin{figure}[t]
% \centering
% \includegraphics[width=9cm,height=7cm]{lirmm.png} 
% \caption{Imaginary vision of BOSS and LIRMM2 images  distribution on a Hyperplane.}
% \label{fig:LIRMM-BOSS}
% \end{figure}

One can observe that results are worst when using {\it BOSS+LIRMM}, compared to only using {\it BOSSBase}. There is 2\% increase of the detection error probabilities for both WOW and S-UNIWARD. For this setup, the enrichment of the learning set is not strongly unbalanced (1 BOSS pair for 2 LIRMM pairs), done with images acquired with different cameras but processed with the same development. {\bf It seems that for a beneficial enrichment, the additional images have to be acquired with the same cameras}. Additional facts seem to confirm this hypothesis in Section \ref{sec:different_sources_unbalance} and Section \ref{sec:RAWimages_different_dev}. 

When enriching the {\it BOSSBase} with {\it BOSS + LIRMM + BOWS2}, the results are as good (or a slightly better for WOW) as using the {\it BOSSBase} alone. Finally, the results become better when {\it BOSS+BOWS2+LIRMM2+VA} is used, but the increase in performance is only of 0.9\% for S-UNIWARD (resp. 2\% for WOW), while using the {\it BOSS+BOWS2} (see Tab.\ref{tab:base_augmentation}) give 2\% increasing for S-UNIWARD (resp. 4\% for WOW). %%%%

Those results confirm again that performance is increased if there is an enrichment with images acquired with the same cameras and with the same development (BOWS-2 share similar cameras and a similar development). This tendency seems to contradict the idea that using millions of images, whose distribution is diverse, would be the best solution for increasing the steganalysis results \cite{Zeng2017_Millions}. Indeed, the added images have to share a very similar "distribution" and images have probably to be acquired with the same cameras. In Section \ref{sec:different_sources_unbalance} we explore a little bit more this hypothesis.

%%%%%%%%%%%%%%%%%%%%%%%%%%%%%%%%%%%%%%%%%%%%%%%%%%%%%%%%%%%%%%%%%%%%%%%%%%%%%%%%
\subsection{Setup 3: Enrichment with strongly dissimilar sources and unbalance proportions}
\label{sec:different_sources_unbalance}

\begin{table}[htb]
\centering
\renewcommand{\arraystretch}{1.6}
\scalebox{0.8}{
\begin{tabular}{l|l|l|l|l|l|}
\cline{2-3}
                                                    
 & \cellcolor[HTML]{C0C0C0}WOW 0.2 bpp 
& \cellcolor[HTML]{C0C0C0}S-UNIWARD 0.2 bpp      \\ \hline                                     
\multicolumn{1}{|l|}{\cellcolor[HTML]{C0C0C0}BOSS}   & {\bf 27.8} \%                        & {\bf 36.7} \%                                                         \\ \hline
\multicolumn{1}{|l|}{\cellcolor[HTML]{C0C0C0}BOSS+PLACES2 1\%} & 34.2 \%                        & 41.6 \%                                         \\ \hline
\multicolumn{1}{|l|}{\cellcolor[HTML]{C0C0C0}BOSS+PLACES2 10\%} & 40.0 \%                        & 43.9 \%                                                         \\ \hline
\multicolumn{1}{|l|}{\cellcolor[HTML]{C0C0C0}BOSS+PLACES2 100\%} & 44.6 \%                        &  45.3 \%                                                                     \\ \hline 
\end{tabular}}
\vspace{+0,4cm}
\caption{Table 4: Base Augmentation influence: error probability of Yedroudj-Net, on WOW and S-UNIWARD at 0.2 bpp with a learning base augmented with different portions of PLACES2. \label{tab:different_sources_unbalance}}
\vspace{+0,5cm}
\end{table}
% \begin{figure}[]
% \centering
% \includegraphics[width=8cm,height=7cm]{Places.png} 
% \caption{Imaginary vision of BOSS and PLACES2 images  distribution on a Hyperplane.}
% \label{fig:BOSS-PLACES}
% \end{figure}

In Table \ref{tab:different_sources_unbalance}, we report the results with {\it no enrichment} (noted {\bf BOSS}), the results with 1\% of PLACES2 enrichment (noted {\bf BOSS + PLACES2 1\%}), the results with 10\% of PLACES2 enrichment (noted {\bf BOSS + PLACES2 10\%}), and 1\% of PLACES2 enrichment (noted {\bf BOSS + PLACES2 100\%}). Note that for {\it PLACES2 1\%}, the training set is made of 14 000 pairs (without counting the validation), for {\it PLACES2 10\%}, the training set is made of 104 000 pairs (without counting the validation), and for the {\it PLACES2 100\%}, the training set is made of 1 004 000 pairs (without counting the validation).

Whatever the enrichment and whatever the embedding algorithm, the results are always worse than using the BOSSBase alone. For the setup where 1\%, resp. 10\%, resp. 100\% of PLACES2 are added to the learning, the results get worse and worse, with respectively an increase of the detection error for S-UNIWARD (resp. WOW) of 5\% (resp. 6\%), 7\% (resp. 12\%), and then 9\% (resp. 17\%). Note that with an enrichment of 100\% of PLACES2 (1 BOSS pair for 251 PLACES2 pairs), the detection is close to a random guessing. 

Since the distribution of BOSS and PLACES2 are totally different (PLACES2 results from a JPEG dequantization, and a very diverse set of sources of cameras), the BOSS distribution is lost, and since no re-balancing measures are used during the learning, the BOSS distribution is considered as anecdotal and it is not really taken into account during the learning. Practically, the total loss computed for BOSS images is negligible compared to the total loss computed for PLACES2 images, and thus a minimization of the global loss will mainly concentrate on minimizing the loss associated to the PLACES2 images. %The training loss was almost 0\% when the validation loss was higher than 1.5\%. Enlevé car les ordres de grandeurs sont inconnus.
Coming back to our previous statement, {\bf using millions of images is not sufficient \cite{Zeng2017_Millions}, the added images have to share a very similar "distribution" and images have probably to be acquired with the same cameras}. 

%A naive vision for this scenario is shown in fig \ref{fig:BOSS-PLACES}. Please note that this figure doesn't represent a real Scatter plot but a hand-made plot. One could conclude that Using a database that does not share properties with the original database is:
% \begin{enumerate}
% \item Inefficient method for data-augmentation to make our CNN achieve better performance.
% \item Create a serious cover-source mismatch phenomenon depends on the number of the added images.
% \end{enumerate}

%%%%%%%%%%%%%%%%%%%%%%%%%%%%%%%%%%%%%%%%%%%%%%%%%%%%%%%%%%%%%%%%%%%%%%%%%%%%%%%%
\subsection{Setup 4: Enrichment with the same RAW images but with a different development}
\label{sec:RAWimages_different_dev}
\begin{table}[htb]
\centering

\renewcommand{\arraystretch}{1.6}
\scalebox{0.8}{
\begin{tabular}{l|l|l|l|l|l|}
\cline{2-3}
                                                    
 & \cellcolor[HTML]{C0C0C0}WOW 0.2 bpp 
& \cellcolor[HTML]{C0C0C0}S-UNIWARD 0.2 bpp      \\ \hline                                     
\multicolumn{1}{|l|}{\cellcolor[HTML]{C0C0C0}BOSS}   & {\bf 27.8} \%                        & {\bf 36.7} \%                                                         \\ \hline
\multicolumn{1}{|l|}{\cellcolor[HTML]{C0C0C0}BOSS+DEV:Res-Bicub} & 25.7 \%                        & 37.5 \%                                         \\ \hline

\multicolumn{1}{|l|}{\cellcolor[HTML]{C0C0C0}BOSS+DEV:Res-Spline} & 26 \%                        & 35.8 \%                                                         \\ \hline
\multicolumn{1}{|l|}{\cellcolor[HTML]{C0C0C0}BOSS+DEV:Res-NoInt} & 25.6 \%                        &36.2  \%                                                                     \\ \hline
\multicolumn{1}{|l|}{\cellcolor[HTML]{C0C0C0}BOSS+DEV:Crop} & 34.8 \%                        &44.2  \%                                                                     \\ \hline
\multicolumn{1}{|l|}{\cellcolor[HTML]{C0C0C0}BOSS+DEV:Res-Crop} & 28.1 \%                        &37.9  \%                                                                     \\ \hline
\multicolumn{1}{|l|}{\cellcolor[HTML]{C0C0C0}BOSS+BOSS-ALP} & 26.0 \%        & 35.5\%                                         \\ \hline

\end{tabular}}
\vspace{+0,4cm}
\caption{Table 5: Base Augmentation influence: error probability of Yedroudj-Net, on WOW and S-UNIWARD at 0.2 bpp with a learning base augmented with different BOSSBase versions.\label{tab:RAWimages_different_dev}}
\vspace{+0,5cm}
\end{table}

In Table \ref{tab:different_sources_unbalance}, we report the results with {\it no enrichment} (noted {\bf BOSS}), and the results with 6 different versions of the BOSSBase, each generated from the RAW images. There is an enrichment with a resizing with a {\it bicubic interpolation} (noted {\bf BOSS+DEV:Res-Bicub}), the an enrichment with a resizing with a {\it spline interpolation} (noted {\bf BOSS+DEV:Res-Spline}), the enrichment with a resizing {\it without any interpolation} (noted {\bf BOSS+DEV:Res-NoInt}), the enrichment with no resizing and a {\it central crop} (noted {\bf BOSS+DEV:Crop}), the enrichment with a resizing to a 768$\times$768 images {\it without any interpolation} and then a {\it central crop} (noted {\bf BOSS+DEV:Res+Crop}), and finally an enrichment with the use of Adobe Photoshop Lightroom 6 instead of {\it ImageMagick}, for generating the color images and then resizing to $256\times256$ the images while keeping the width/length ratio (noted {\bf BOSS-APL}).

From Table \ref{tab:RAWimages_different_dev}, we can observe that the enrichment with a {\it crop development} (BOSS+DEV:Crop) lead to very bad results. The increase of the detection error of 7\% for S-UNIWARD (resp. 7\% for WOW).  The enrichment with a resize to 768$\times$768 followed by a crop (BOSS+DEV:Res+Crop), to a lesser extent, also give bad results with an increase of the detection error of 1\% for S-UNIWARD (resp. 0.3\% for WOW). Those bad results suggest that a resolution change during the development has a strong impact on the pixels distributions. When looking to the extreme case of the {\it crop development} (BOSS+DEV:Crop), we easily understand that the resulting images content change; there is almost no variations and no edges. Thus, {\bf an enrichment with a BOSS version whose development does not ensure the same final pixel resolution than BOSS Base will not enrich favourably the learning data-base}.

In counterpart, using the same resize procedure with a slight variation on the {\it interpolation} (spline, no-interpolation, bicubic), or with the {\it Adobe Photoshop Lightroom Process} allows scrounging at most 1\% for S-UNIWARD (resp. 2\% for WOW). This confirms that additional samples very close to the target BOSS distribution can improve the learning capabilities. {\bf Looking back to the various experiment done previously, one can observe that in order to enrich favourably a target database, a favourable measure is to use images acquired with the same cameras than the target database, and to use a very close resizing process than the one used for the target database.}

\begin{table}[htb]
\centering
\renewcommand{\arraystretch}{1.6}
\scalebox{0.8}{
\begin{tabular}{l|l|l|l|l|l|}
\cline{2-3}
                                                    
 & \cellcolor[HTML]{C0C0C0}WOW 0.2 bpp 
& \cellcolor[HTML]{C0C0C0}S-UNIWARD 0.2 bpp      \\ \hline                                     
\multicolumn{1}{|l|}{\cellcolor[HTML]{C0C0C0}BOSS}   & {\bf 27.8} \%                        & {\bf 36.7} \%                                                         \\ \hline
\multicolumn{1}{|l|}{\cellcolor[HTML]{C0C0C0}BOSS+all-DEV} & 23.0 \%                        & 33.2 \%                                         \\ \hline
\multicolumn{1}{|l|}{\cellcolor[HTML]{C0C0C0}BOSS+BOWS2} & 23.7 \%        & 34.4\%                                         \\ \hline
\end{tabular}}
\vspace{+0,4cm}
\caption{Table 6: Base Augmentation influence: error probability of  Yedroudj-Net, on WOW and S-UNIWARD at 0.2 bpp with a learning base augmented with different versions of BOSSBase.\label{tab:RAWimages_all_dev}}
\vspace{+0,5cm}
\end{table}

In order to push the reflection a little bit more, we made an additional experiment where we regrouped diverse versions of BOSSBase (BOSS+DEV:Res+Bicub, BOSS+DEV:Res+Spline, BOSS+DEV:Res+NoInt, BOSS+DEV:Res+Crop) to the exception of BOSS+DEV:Crop. In Table \ref{tab:RAWimages_all_dev}, we report the results with this gathering of various development (noted {\bf BOSS+all-DEV}), and the results with LIRMM and BOWS2 enrichment (noted {\bf LIRMM+BOWS2} and already reported in Section \ref{sec:classical_enrichment}). Note that for {\it BOSS+all-DEV}, the training set is made of 44 000 pairs (without counting the validation), and for {\it LIRMM+BOWS2} the training set is made of 14 000 pairs (without counting the validation).

For those two enrichments, there is a real improvement with a decrease of the error probability of detection of 2-3\% for S-UNIWARD (and 4\% for WOW). This last result is very interesting and shows that in order to enrich a database, in a practical scenario, there are at least those two options: 
%\shadowbox{\\Given a target database: }
\colorbox{lightgray}{
  \parbox{8cm}{
Given a target database:
\begin{itemize}
\item either Eve (the steganalyst) finds the same camera(s) (used for generating the target database), capture new images, and reproduce the same development than the target database, with a special caution to the resizing, 
\item either Eve has an access to the original RAW images and reproduce similar developments than the target database with the similar resizing, 
\end{itemize}
The reader should also remember that the Virtual Augmentation is also a good cheap processing measure.
}}

Note that it is unclear which option would be better in a practical case. Additional experiments have to be done in the future. Anyway, those two enrichments show that a very caution process has to be taken for really improving the results. We believe that those enrichments reduce the over-fitting and also improve the generalization of the learner. %We also believe CNN should reinforce this measure (? -> Loss à changer comme dans les GAN ?).

\subsection{Setup 5: Enrichment with a re-development of the learning set}
\label{sec:images_redevelop}

In all previous setups, given a target database (never compressed 8-bits grey-level 256$\times$256 images), we were presuming either a prior knowledge of the cameras used for the images acquisitions or a direct access to the RAW versions of the original images. In real-world cases, those knowledges are most of the time not available. Moreover, retrieving the camera models is a very complicated task in a real scenario due to the huge number of cameras. 

\begin{table}[htb]
\centering
\renewcommand{\arraystretch}{1.6}
\scalebox{0.8}{
\begin{tabular}{l|l|l|l|l|l|}
\cline{2-3}
                                                    
 & \cellcolor[HTML]{C0C0C0}WOW 0.2 bpp 
& \cellcolor[HTML]{C0C0C0}S-UNIWARD 0.2 bpp      \\ \hline                                     
\multicolumn{1}{|l|}{\cellcolor[HTML]{C0C0C0}BOSS}   & {\bf 27.8} \%                        & {\bf 36.7} \%                                                         \\ \hline
\multicolumn{1}{|l|}{\cellcolor[HTML]{C0C0C0}BOSS+DEV:Translation} & 34.7.0 \%                        & 47.8 \%                                         \\ \hline
\multicolumn{1}{|l|}{\cellcolor[HTML]{C0C0C0}BOSS+DEV:Up-Down-Sampling} & 31.2 \%                        & 42.6 \%                                         \\ \hline

\end{tabular}}
\vspace{+0,4cm}
\caption{Table 7: Base Augmentation influence: error probability of Yedroudj-Net, on WOW and S-UNIWARD at 0.2 bpp with a learning base augmented with a re-development of BOSSBase.\label{tab:images_redevelop}}
\vspace{+0,5cm}
\end{table}

In Table \ref{tab:images_redevelop}, we report the results with {\it no enrichment} (noted {\bf BOSS}), and the results with 2 different redeveloped versions of the BOSSBase, each generated from the original 256$\times$256 8-bits grey-level BOSSBase images. The first redevelopment (noted {\bf BOSS+DEV:Translation}) consists in applying a sub-pixel image translation, of 0.5 pixel, on the padded (symmetric padding) images, and then applying a crop operation to re-obtain a 256$\times$256 images. The second redevelopment (noted {\bf BOSS+DEV:Up-Down-Sampling}) consists in applying a Lanczos3 filter for the up-sampling in order to obtain a 512$\times$512 images, and then down-sampling with the same interpolation Kernel to re-obtain images of 256$\times$256 size. The results are catastrophic with an increase of the error probability of 6\% to 11\% for S-UNIWARD and 4\% to 7\% for WOW. The use of a redevelopment does not seem to be a good idea.

% \section{Discussion}
% Faut il une discussion ?
% \begin{itemize}
% \item Rappel de la conclusion ?
% \item Discussion du positionnement par rapport à la notion de source-mismatch,
% \item discussion sur les mesures intéressantes ? Partitionnement prealable (comme prop EUSIPCO J. Pasquet) ? Recherche de cameara (?), Apprendre sur des millions d'exemples (ref Ker + Jiwu), Ensemble ?, Transfer ?, Domain Transfert ?, GAN, regularisation des Loss ?
% \end{itemize}

\section{Conclusion}
In this paper, we have explored ways to enrich a learning database when steganalysis is done with a CNN. The enrichment is a crucial task since, in the majority of the today's experiments, the required number of images have to be extremely high due to the huge number of parameters to be learned. Using an insufficient set of examples (images) leads to CNNs that have not "learned enough" and the average efficiency is thus reduced.

After recalling the state-of-the-art of 2017 for the spatial CNN steganalysis, and briefly recalling the state-of-the-art steganalysis approach named Yedroudj-Net, we have presented various results. Additionally to the classical data augmentation which consists to apply flips and rotations on the learning images \cite{Krizhevsky_AlexNet_2012}, we observed two others ways for favorably enriching the learning database. The trend is that, in a clairvoyant scenario (knowledge of the embedding algorithm, knowledge of the payload size, approximate knowledge of the of the images distribution), for a given target (test) database,  in order to augment its learning database, the steganalyst (Eve) has two choices: 
\begin{itemize}
\item Either she is able to guess the camera(s) used for generating the target database. She thus captures new images, and reproduce a similar development than the target database, with a special caution to the resizing,
\item Either she has an access to the original RAW images and reproduces a similar development than the target database with the similar resizing.
\end{itemize}

Those two possible ways to enrich the database are very restrictive. As explained in the paper some complementary solutions can be used such as transfer learning \cite{Qian2016_Transfer}, or the use of ensembles \cite{Xu2016b}, but the underlying questions of generalizations / cover-source mismatch have to be explored deeper in the future.

\bibliographystyle{IEEEtran}
\bibliography{biblio}

%%%%%%%%%%%%%%%%%%%%%%%%%%%%%%%%%%
% Reference Preparation
%%%%%%%%%%%%%%%%%%%%%%%%%%%%%%%%%%

%\section{Reference Preparation}
%Use the standard LaTeX \emph{cite} command for references in the
%text. You can then use the standard bibliography command to generate
%the list of references. Add the command \emph{small} before the
%bibliography to give it the right font size.  Reference \cite{Qian_2015_Deep}
%style should be used for books, Reference \cite{Pibre2016} style should be
%used for Journals, and Reference \cite{Xu2016b} style should be used for
%Proceedings.

\section*{Acknowledgments} 
This work was supported by the University of Montpellier (LIRMM), and the Algerian Ministry of Higher Education / Scientific Research.


%%%%%%%%%%%%%%%%%%%%%%%%%%%%%%%%%%
% Bibliography
%%%%%%%%%%%%%%%%%%%%%%%%%%%%%%%%%%

\begin{biography}

\hspace{0.5cm} Mehdi YEDROUDJ attained his Master's degree in Computer Science in 2016 from the University of Constantine 2, Algeria. He is currently working toward the Ph.D. degree in the LIRMM laboratory of Montpellier. His research interests are steganography/steganalysis.

Fr\'ed\'eric COMBY received his M.Sc. degree in automatic
and micro-electronic systems in 1998, and the Ph.D. degree in automatic and signal
processing in 2001 from the University of Montpellier, France. 
He joined the ICAR Team (image and interaction), in the LIRMM (Laboratory
of Informatics, Robotics, and Microelectronics of Montpellier) as Assistant
Professor in 2003. His current research topics include image processing, 
vision and multimedia security.

Marc CHAUMONT received his Engineer Diploma in Computer Sciences at the INSA 
(National Institute of Applied Sciences) of Rennes, France in 1999, his Ph.D. 
in Computer Sciences at the IRISA Rennes (INRIA, CNRS, University of Rennes 2, and INSA) 
in 2003, and his HDR ("Habilitation \`a Diriger des Recherches") at the University of Montpellier 
in 2013. Since September 2005, he is an Assistant Professor in the LIRMM laboratory
(Laboratory of Computer Science, Robotics and Microelectronic) of Montpellier and the University of N\^imes. 
His main research interests are in steganography, steganalysis, digital image forensic, and objects detection with Deep Learning. 
He is a retired member of the TC of IEEE SPS - Information Forensics and Security, and a reviewer for more than 20 journals (IEEE TIFS, IS\&T JEI, ...) and for more than 10 conferences (EI MWSF, IEEE WIFS, ACM IH\&MMSec, IEEE ICIP, ...).

\end{biography}

\end{document}